%% file: fmodes.tex
\documentclass[prd, aps, nofootinbib, preprintnumbers, showpacs, superscriptaddress, twocolumn, floatfix]{revtex4}

\usepackage{amssymb}
\usepackage[english]{babel}
\usepackage{ifpdf}
\usepackage[latin9]{inputenc}
\usepackage{mathpazo}
\usepackage{mathrsfs}
\usepackage{float}
\usepackage{psfrag}
\usepackage{color}

\ifpdf
\usepackage{hyperref}
\fi

\usepackage{graphicx}

\newcommand{\be}{\begin{equation}}
\newcommand{\ee}{\end{equation}}

\newcommand{\bee}{\begin{eqnarray}}
\newcommand{\eee}{\end{eqnarray}}

\begin{document}

\title{On the frequency band of the $f$-mode CFS instability}

\pacs{04.30.Db,    
         04.40.Dg,     
         95.30.Sf,       
         97.10.Sj 	      
     }

\author{Burkhard~Zink} 
\affiliation{Theoretical Astrophysics, University of T\"ubingen, Auf der Morgenstelle 10, 
    T\"ubingen 72076, Germany}

\author{Oleg~Korobkin}
\affiliation{Center for Computation and
Technology, Louisiana State University, Baton Rouge, LA 70803, USA}

\author{Erik~Schnetter}
\affiliation{Center for Computation and
Technology, Louisiana State University, Baton Rouge, LA 70803, USA}
\affiliation{Department of Physics \& Astronomy, Louisiana State University, Baton Rouge, LA 70803, USA}

\author{Nikolaos~Stergioulas}
\affiliation{Department of Physics, Section of Astrophysics, Astronomy and Mechanics
Aristotle, University of Thessaloniki, Thessaloniki, 54124 Greece}

\begin{abstract}

Rapidly rotating neutron stars can be unstable to the gravitational-wave-driven CFS mechanism if they have a neutral
point in the spectrum of nonaxisymmetric $f$-modes. We investigate the frequencies of these
modes in two sequences of uniformly rotating polytropes using nonlinear 
simulations in full general relativity, determine the approximate locations of the neutral points,
and derive limits on the observable frequency band available to the instability in these sequences.
We find that general relativity enhances the detectability of a CFS-unstable neutron star substantially,
both by widening the instability window and enlarging the band into the optimal range for interferometric 
detectors like LIGO, VIRGO, and GEO-600.

\end{abstract}

\maketitle

\input{introduction}

\input{techniques}

\input{tests}

\input{results}

\input{conclusion}

\acknowledgments{The authors would like to thank G. Corvino, K. Kokkotas and S. Yoshida
for helpful comments and discussions.
This work was supported by the Sonderforschungsbereich/Transregio 7
on gravitational wave astronomy by the DPG, the NSF TeraGrid, the Louisiana Optical Network
Initiative (LONI), the Albert Einstein Institute, and the University of T\"ubingen. Computations
were performed on the LONI cluster \emph{queenbee}, the TeraGrid clusters \emph{ranger}
and \emph{kraken} on allocation TG-MCA02N014, and the \emph{damiana} cluster at the AEI. The code used in this study
is based on the \emph{Cactus} computational infrastructure, and the \emph{Carpet}
mesh refinement driver. The development of the multipatch infrastructure was supported
by NSF SDCI 0721915 and NSF PIF 0701566. B.Z. thankfully acknowledges support from the AEI and L. Rezzolla.}


\end{document}

%% file: introduction.tex
\section{Introduction}
\label{sec:introduction}

The possibility of detecting gravitational waves from neutron stars is an exciting prospect for present and planned observatories,
since it may be our only direct probe into the very core of the star (neutrinos being another possibility). In addition,
neutron stars have a very unique spectrum of stable and unstable modes of oscillation which could be
identified in gravitational wave observations, and indirect conclusions concerning the equation of state and internal structure of the
star could be drawn similar to helio- and asteroseismology.

Instabilities are particularly interesting in this context, since in this case a spectral line can be amplified either directly (by dynamical transition)
or driven by radiative fields to potentially large saturation amplitudes, thereby enhancing the chances of detection. A number of
neutron star instabilities are known (for a review see \cite{Stergioulas03}) which will be active subject to particular (different) requirements
on the stellar characteristics. The resulting gravitational wave signals are expected to carry signatures unique to the unstable mode and 
the structure of the star.

Typical pressure oscillations in neutron stars have frequencies in the kHz band, and are 
therefore, even if unstable, not located in the optimal range of sensitivity
of interferometric detectors like LIGO, VIRGO and GEO-600. However,
a particular class of modes, the nonaxisymmetric $f$-modes, also have a counter-rotating branch which extends to very
low frequencies in the range of hundreds of Hertz, which is optimal for detection by laser interferometers. These modes are
the subject of the current study.

These counter-rotating $f$-modes could grow to large amplitudes (and produce correspondingly larger gravitational-wave strains)
in a manner first discovered by Chandrasekhar, Friedman and Schutz \cite{Chandrasekhar70a, Friedman75a}, in which
the dynamical gravitational field itself is the agent driving the instability. This so-called CFS mechanism operates when
a mode that is counter-rotating with respect to the star appears as co-rotating in the inertial frame, its frequency having
crossed the \emph{neutral point} (zero frequency in the inertial frame) due to the high rotation rate of the star. 
If the neutron star spins faster than critical the instability could be active, with some qualifications we will mention below.

The most important one of these unstable modes is likely the lowest-order one with spherical harmonic indices $\ell = m = 2$.
In Newtonian gravity, this mode becomes unstable at spin rates which are too high for typical models of (uniformly rotating)
neutron stars. General relativity enhances the instability substantially, as demonstrated first by Stergioulas and
Friedman \cite{Stergioulas98a}, and therefore makes it a more promising source for detection.

We will focus here on $f$-modes, which have pressure gradients
as the restoring force. Since no successful linear method exists to address this problem in a fully relativistic
stellar model, we employ full nonlinear numerical simulations to extract the mode frequencies.

The CFS mechanism is not restricted to this class of modes. The only requirement is that, along a sequence of stars
with increasing rotational over binding energy $T/|W|$, a mode that is counter-rotating in the rest frame of the star changes type
to corotation in the frame of a faraway observer. Gravitational waves extract positive angular momentum from
the star, and the mode's angular momentum becomes increasingly more negative with respect to the star, 
i.e. its amplitude increases with time. $r$-modes (a class of inertial modes)
and $w$-modes (trapped spacetime oscillations) have been discussed in this context as well \cite{Andersson98a, Kokkotas04}.

The actual instability window depends on a number of parameters. The lower limit in terms of rotation is given by the neutral point,
which for the $\ell = m = 2$ bar-mode in Newtonian gravity has been found to be close to $T/|W| = 0.14$. Higher
order modes with $\ell = m$ have neutral points at \emph{lower} $T/|W|$, i.e. they have a larger instability window. But the growth 
timescale of the instability also depends on $m$, such that higher order modes need a longer time to grow to high amplitudes. In
that, the instability is competing with viscosity in the star which damps the mode amplitude. Instability requires 
$1/\tau = 1/\tau_{\mathrm{GW}} + 1/\tau_{\mathrm{viscosity}}$ to be positive, where $\tau$ denotes the growth timescale of the particular mechanism.
In particular, shear and bulk viscosity in the neutron star provide damping agents which leave an effective temperature
window around $10^{8} \ldots 10^{11} K$ and mode numbers $m \leq 5$ \cite{Ipser91}.

A superfluid core in the neutron star may also be instrumental in suppressing the $f$-mode CFS instability by mutual friction between
quantized neutron vortices \cite{Lindblom95, Andersson09}. This can only occur below the transition temperature to a
superfluid state ($\approx 10^9 K$) and is therefore only relevant for older neutron stars. The effectiveness of mutual friction
is based on estimates for the coupling strength. Also, current predictions for the instability windows assume Newtonian physics,
while general relativity is known to enhance the gravitational instability. Therefore it is not yet clear whether the instability
could still be active also in some old neutron stars (the $r$-mode instability could also provide the dominant spin-limit
mechanism in those cases).

Another unknown is the saturation amplitude of the deformation. In the case of the CFS instability for the $r$-modes, previous
studies have indicated that nonlinear mode-mode couplings appear to be effective in saturating the instability at low 
amplitudes \cite{Gressman02a, Arras03, Brink04a, Bondarescu07a}. A similar effect could occur for the $f$-mode instability, and whether this is the case or not is presently
unknown. Since the spectrum of pressure modes is less dense than the inertial mode spectrum and has
a different characteristic eigenfunction, nonlinear mode-mode couplings could be less effective in this case. Also, some numerical
evidence for the development of the CFS instability in rapidly rotating Newtonian polytropes \cite{Shibata04b, Shangli04a} seems to 
indicate that large amplitudes could be reached (but these studies also enhance the growth rate by a large factor, so
they may not represent the actual mode coupling accurately). If that is the case, estimates using equilibrium ellipsoids \cite{Lai95} 
predict that the gravitational wave signal would sweep through a band in the range of 100's of Hz and could be visible
up to 140 Mpc in Advanced LIGO.

The $f$-mode CFS instability requires a $T/|W|$ of several percent, which is low compared to the dynamical bar-mode 
($T/|W| \approx 0.24$) 
but still rather high compared to typical estimated iron core rotation rates. On the other hand, a subset of massive stars may undergo an evolution in which
angular momentum transport from the core is less efficient than in typical cases, which is consistent with (but not supported by)
the collapsar model for long gamma-ray bursts which requires a dense disk of sufficient angular momentum as a source for
powering the engine. To have a reasonable chance of detecting a CFS-unstable star, it is therefore important to reach
high signal-to-noise ratios for a sufficient event rate.

This study is focused on the frequency band and the location of the neutral point for the $m = 2$ and $m = 3$ modes.
The paper is organized as follows: in Section~\ref{sec:techniques}, we describe the physical model,
our numerical techniques, and the equilibrium stars used as initial data for time evolution. In Section~\ref{sec:tests},
we present results from code tests and compare to previous work. Section~\ref{sec:results} contains the results of
our study, and in Section~\ref{sec:conclusion} we conclude with a discussion.

%% file: techniques.tex
\section{Physical Model and Numerical Techniques}
\label{sec:techniques}

\subsection{Basic physical model}

As mentioned in the introduction, we model neutron stars as (uniformly rotating) polytropes, i.e. those equilibrium stars whose structure follows from the pressure-density relation $P = K \, \rho^\Gamma$, with $P$ the pressure, $\rho$ the rest-mass density, and $K$ and $\Gamma$ free parameters. This model is simplified in a number of ways. Actual neutron stars are expected to consist of a composition of baryons and leptons which depends on the local thermodynamical properties and the evolutionary history of the star \cite{Glendenning:1996}. In addition to protons and neutrons, the core of the neutron star may also contain hyperons or kaons, and it is possible that a deconfined equilibrium quark phase exists at high densities. These uncertainties in the microstructure affect the macroscopic structure and dynamics as well, in particular since different compositions have a different stiffness in reaction to perturbations. However, polytropes have the advantage to essentially only depend on one free parameter (the constant $K$ can be viewed as a scaling parameter, since different choices of $K$ will lead to the same dynamics in a properly scaled system of units, see also the discussion in \cite{Stergioulas98a}), and since they have been used in many other studies they make it easier to isolate effects not associated with the microstructure, e.g. the choice of Newtonian vs. relativistic models.

For the time development of perturbations we also assume the star to be an ideal fluid governed by the gamma-law $P = (\Gamma - 1) \, \rho \, \epsilon$, where we select $\Gamma$ to be the same as in the polytropic relation, and $\epsilon$ is the specific internal energy density of a fluid element as measured in a local rest frame. In addition, for numerical reasons, we restrict attention to purely adiabatic perturbations, i.e. we neglect local heat exchange by projecting the pressure back to $P = K \, \rho^\Gamma$ -- this procedure is commonly referred to as assuming a \emph{polytropic} equation of state. It is generally used to reduce artefacts associated with evolving the surface of the polytrope.

In particular, these assumptions neglect non-ideal effects like superfluidity \cite{Andersson:2007a} and solid crusts \cite{Chamel:2008a}, which are important in neutron stars below $10^{9} \ldots 10^{10} \, \mathrm{K}$, i.e. likely already a short time after birth. We also neglect magnetic fields by assumption of a pure ideal fluid. The resulting model is therefore not suited to give quantitatively correct results, but allows us to qualitatively discuss the effect of including full general relativity as part of the model, specifically in comparison to earlier results in the Cowling approximation \cite{Gaertig08a, Krueger09a}.

\subsection{Numerical method}

We study neutron star oscillations by the time development of perturbed equilibrium stars in full general relativistic hydrodynamics. The numerical code used has been developed inside the \emph{Cactus} computational infrastructure \cite{Goodale02a}, and uses the \emph{Carpet} mesh refinement and multi-mesh driver \cite{Schnetter-etal-03b}. The module for evolving the general relativistic fluid fields is the \emph{Typhon} multi-block code \cite{Zink:2008b}, which has been coupled to a multi-block implementation of the generalized harmonic form of Einstein's field equations \cite{Lindblom:2005qh, Schnetter06a-nourl}. We point the reader to these publications concerning the details of the differential equations and the discrete implementation.

The generalized harmonic evolution code provides a linearly stable initial value problem by virtue of being a first-order symmetric hyperbolic system. In addition, we use maximally dissipative outer boundary conditions discretized with linearly stable penalty operators, and preserve the linear stability in a discrete sense by using one-sided finite difference operators satisfying the summation-by-parts property (see \cite{Lindblom:2005qh, Schnetter06a-nourl} and references). The scheme also employs a constraint-damping technique and Kreiss-Oliger type dissipation operators of high order (we usually employ 8th-order accurate finite difference operators in the interior) to stabilize the non-linear system. These properties are particularly important for long-term stable evolutions of black holes and neutron stars. In particular, we are able to evolve isolated neutron stars for apparently arbitrary long times with domain boundaries very close to the stellar surface, and see only minimal growth of constraint violations. The accuracy of the numerical evolution system is nowadays clearly dominated by the (low-order accurate) methods for evolving the fluid fields.

For the fluid fields, we employ a standard finite-volume discretization with PPM reconstruction and HLL flux estimates. The evolution systems are coupled via the method of lines, and the time evolution is performed with 3rd-order Runge-Kutta integrators. Since the fluid scheme is unable to deal with matter-vacuum interfaces, we employ a low-density artificial atmosphere outside of the isolated star to regularize the method.

Initial data for the time evolution is obtained with the \emph{rns} code \cite{Stergioulas95} and mapped to the numerical domain. Directly afterwards, a perturbation of the fluid fields is added to the equilibrium star to preferentially excite the relevant oscillation modes. Mode frequency extraction is performed by either extracting time profiles of fluid quantities at certain locations inside the star (in particular the center), or performing discrete Fourier transforms on coordinate circles in the equatorial plane \cite{Tohline85, Zink2006a}. The resulting time evolutions are then again Fourier-transformed in post-processing to obtain power spectra.

\subsection{Initial models}
\label{sec:initial_models}

As mentioned earlier, all initial models used in this paper have been generated with the \emph{rns} code. A model is specified by its central density $\rho_c$, the ratio of polar to equatorial coordinate radii, and the polytropic constants $K$ and $\Gamma$. In addition, the resolution of the two-dimensional grid $n_s, n_m$ can be adjusted, as well as the target accuracy for the iterative procedure $e_{acc}$, and the maximal spectral coefficient $\ell_{max}$. We fix these values to $n_s = 601$, $n_m = 301$, $e_{acc} = 10^{-8}$ and $\ell_{max} = 10$, which are sufficient to reproduce properties of published models up to about $10^{-3} \ldots 10^{-4}$ relative accuracy.

For purposes of testing the code, we use models from the well-investigated BU sequence \cite{Font99, Font01, Dimmelmeier06a, Gaertig08a}, which is a set of uniformly rotating polytropes with $\Gamma = 2$, $K = 100$ and $\rho_c = 1.28 \times 10^{-3}$.\footnote{A note on units: The numerical code implicitly employs geometrized units $G = c = 1$. Since polytropes are scale invariant, each choice of $K$ needs to be supplemented by an additional choice of length to convert to cgs units. In fact, $K = 1$ would be sufficient, and we use different values only for numerical convenience. Any sequence of polytropes gives results for all possible models of the same stratification, i.e. the range of masses in terms of cgs units can be fixed by \emph{observational} constraints. (A more precise analysis would make little sense here, given that polytropes are a substantial simplification of the actual interior structure of neutron stars.) For the sequence BU test models, we will fix the length scale to $M_\odot = 1$, which implies $\rho_c = 1.28 \times 10^{-3} c^6 / G^3 M_\odot^2 \approx 7.906 \times 10^{14} \mathrm{g}/\mathrm{cm}^3$ at a mass range $M = 1.4 \ldots 1.7 M_\odot$.} The properties of the nonrotating model BU0 and the rapidly rotating BU7 used for our comparisons are listed in Table~\ref{table:bu_models}. For converting to cgs units later, we will assume $M_\odot = 1$.

\begin{table}
\begin{center}
\begin{tabular}{cccccc}
Model & $\rho_c$ & $r_p/r_e$ & $M$ & $T/|W|$ & $\Omega / \Omega_K$ \\
          &  $\times 10^{-3}$  &                 &     & $\times 10^{-2}$ &  \\
\hline
BU0 & 1.28 & 1.00 & 1.400 & 0.000 & 0.000 \\
BU7 & 1.28 & 0.65 & 1.666 & 8.439 & 0.850 \\
\hline
\end{tabular}
\end{center}
\caption{Properties of the models BU0 and BU7 used for comparison with previous publications. The
columns are the central density $\rho_c$, ratio of polar over equatorial coordinate radii $r_p/r_e$,
gravitational mass $M$, ratio of kinetic over gravitational binding energy $T/|W|$, and ratio
of angular velocity over the Kepler limit $\Omega / \Omega_K$. All models are uniformly rotating polytropes
with $K = 100$ and $\Gamma = 2$.}
\label{table:bu_models}
\end{table}

In addition to these models, we have constructed a sequence of $\Gamma = 2$ polytropes with a higher central density $\rho_c = 2.6812 \times 10^{-3}$, for purposes of comparing with the approximate estimates of the location of $f$-mode neutral points in full general relativity by Stergioulas \& Friedman \cite{Stergioulas98a}. This choice results in a central energy density of $e_c = 3.4 \times 10^{-3}$. The properties of selected models of this ``S'' sequence are listed in Table~\ref{table:s_models}. Finally, to investigate the effects of stiffness in the equation of state, we also define a sequence with $\Gamma = 2.5$, which we will denote ``C'' sequence. Selected members are listed in Table~\ref{table:c_models}. Note that the fastest rotating member, C3, could only be constructed by lowering the convergence factor of \emph{rns} to $0.3$. Since the mass of the nonrotating member is $M (\Omega = 0) = 1.3056$, we convert to cgs units with a length factor $L_{cgs} = 1.9226 \times 10^{5} \mathrm{cm}$: this results in a mass of the nonrotating neutron star of $M(\Omega = 0) = 1.7 M_\odot$, which is more consistent with expectations from observations. We repeat that this choice is arbitrary, and our results for frequencies can be rescaled to any other mass range as desired.

\begin{table}
\begin{center}
\begin{tabular}{cccccc}
Model & $\rho_c$ & $r_p/r_e$ & $M$ & $T/|W|$ & $\Omega / \Omega_K$ \\
          &   $\times 10^{-3}$  &                 &     & $\times 10^{-2}$ &  \\
\hline
S0 & 2.6812 & 1.0000 & 1.6287 & 0.000 & 0.000 \\
S1 & 2.6812 & 0.9160 & 1.6801 & 2.001 & 0.380 \\
S2 & 2.6812 & 0.8338 & 1.7367 & 4.001 & 0.552 \\
S3 & 2.6812 & 0.7487 & 1.7995 & 6.001 & 0.704 \\
S4 & 2.6812 & 0.6393 & 1.8700 & 8.000 & 0.894 \\
\hline
\end{tabular}
\end{center}
\caption{Properties of the models S0 to S4 used for comparison of the neutral point with Stergioulas \& Friedman \cite{Stergioulas98a}. All
models are uniformly rotating polytropes with $K = 100$ and $\Gamma = 2$. The columns are the same as in Table~\ref{table:bu_models}. The 
particular members have been selected to give uniform spacing in $T/|W|$.}
\label{table:s_models}
\end{table}

\begin{table}
\begin{center}
\begin{tabular}{cccccc}
Model & $\rho_c$ & $r_p/r_e$ & $M$ & $T/|W|$ & $\Omega / \Omega_K$ \\
          &   $\times 10^{-3}$  &                 &     & $\times 10^{-2}$ &  \\
\hline
C0 & 5.0 & 1.0000 & 1.3056 & 0.000 & 0.000 \\
C1 & 5.0 & 0.8699 & 1.3598 & 4.000 & 0.472 \\
C2 & 5.0 & 0.7467 & 1.4499 & 8.000 & 0.686 \\
C3 & 5.0 & 0.6040 & 1.5609 & 12.00 & 0.910 \\
\hline
\end{tabular}
\end{center}
\caption{Properties of the models C0 to C3 used for investigating the effects of stiffness on nonaxisymmetric $f$-modes. All
models are uniformly rotating polytropes with $K = 1000$ and $\Gamma = 2.5$. The columns are the same as in Table~\ref{table:bu_models}.}
\label{table:c_models}
\end{table}

%% file: tests.tex
\section{Code Tests}
\label{sec:tests}

In this section, we present results from a number of standard code tests and compare to published results for neutron star oscillations. For the case of stationary stars in the Cowling approximation without artificial perturbation, but covered by multiple blocks, results have already been presented in \cite{Zink2007a}. In this study, we will not need the additional complexity of multi-block systems and instead resort to uniform grids to save computational cost.

\subsection{Oscillations of a nonrotating neutron star}

Oscillation frequencies in nonrotating general relativistic polytropes, and in particular in the BU0 model which we consider here, have been reported in \cite{Font01b, Font01, Gaertig08a} in the Cowling approximation (i.e. keeping the spacetime fixed) and in \cite{ Dimmelmeier06a} in general relativity subject to the conformal flatness condtion. To compare with these results, we have generated the model with \emph{rns} and mapped it to a uniform grid with coordinate extent $[-13,13]^3$. The stellar surface is located at coordinate radius $R = 8.2$, which means that the outer boundary is at $\approx 1.6$ times the stellar radius. This ratio is sufficient to avoid artefacts from surface-outer boundary interactions in the fluid system. The techniques used for the spacetime evolution, in particular the use of maximally dissipative outer boundary conditions via penalty operators\footnote{In principle, constraint-preserving outer boundary conditions \cite{Calabrese02d, Seiler08a} could be a superior choice, but we find that for studies of fluid modes maximally dissipative boundary conditions give good enough results.} and the constraint-damping property of the generalized harmonic evolution enforce stability of the spacetime system and only a moderate growth of constraint violations. Experiments with larger coordinate domains (while keeping the resolution fixed) produce no difference in the resulting fluid mode frequencies within the given error bars.

\begin{figure}
\includegraphics[width=\columnwidth]{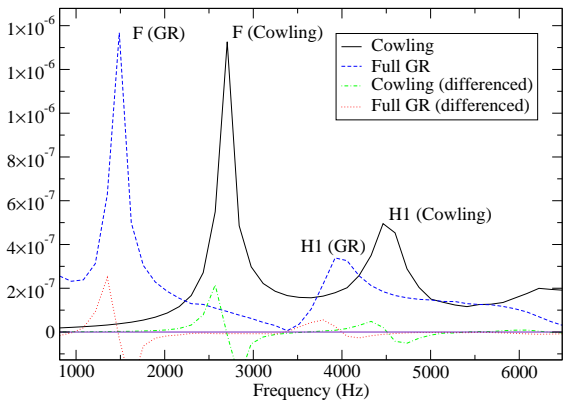}
\caption{Fundamental mode in the spherically symmetric polytrope BU0 ($\Gamma = 2.0$, $K = 100$, central density
$\rho_c = 1.28 \times 10^{-3}$): The mode is excited with a nodeless spherically symmetric density perturbation. This plot
shows a discrete Fourier transform of the central density time evolution $\rho_c (t)$, for both a model in the Cowling
approximation and in full general relativity. The resulting major peak is the fundamental mode F, which can be compared
to published estimates. Since we do not impose an exact eigenfunction, the initial perturbation (and also the discretization error)
also have some overlap with the first overtone H1, although the perturbation is nodeless. The differenced plots refer to central
finite differences of the Fourier transforms, which allow to locate the maxima of the peaks more accurately. (The difference 
graph values have been rescaled for illustration.)}
\label{fig:bu0_F_cowling_vs_gr}
\end{figure}

The spherically symmetric fundamental mode (usually denoted $F$) is already excited by discretization errors, but to be consistent with the remaining study and with published work we impose a simple perturbation of the form \cite{Font01}
\begin{equation}
\delta \rho = A \rho_c \cos \left[\frac{\pi r}{2 r_S} \right]
\end{equation}
where $\delta \rho$ is the perturbation in rest-mass density, $A$ an arbitrary amplitude, $r$ the radial coordinate, and $r_S$ is the coordinate radius of the surface. For this mode it is sufficient to extract the time evolution of the central density and perform a Fourier transform on the profile. Fig.~\ref{fig:bu0_F_cowling_vs_gr} compares the spectra obtained from the Cowling approximation with fully general relativistic models. All data were obtained using a small perturbation amplitude ($A = 10^{-3}$) and from an evolution over $10 \, \mathrm{ms}$. The maximum evolution time we use is not limited by numerical stability, but by numerical dissipation present in the hydrodynamical scheme (this aspect will be discussed in more detail in the context of the nonaxisymmetric mode extraction below). 

Fig.~\ref{fig:bu0_F_cowling_vs_gr} shows that the fundamental mode frequencies are very different for models in the Cowling approximation and fully general relativistic models (see also e.g. \cite{Dimmelmeier06a}). In particular, we obtain a frequency of $\nu(F) = 2694 (40) \, \mathrm{Hz}$ in the Cowling approximation, which is in agreement with \cite{Font01} ($2706 \, \mathrm{Hz}$), and for full general relativity $\nu(F) = 1465 (40) \, \mathrm{Hz}$, compared to $1458 \, \mathrm{Hz}$ in \cite{Dimmelmeier06a}. All error bars are rough estimates based on the distance between the root location in the first derivative of the Fourier spectrum and the next non-zero value.

Non-radial modes in this model have been investigated as well, in particular the set of $\ell = 2$ $f$-modes (which are degenerate in $m$ since the 
background is spherically symmetric). To excite these modes, we impose a perturbation of the form 
\begin{equation}
\delta \rho = A \rho_c \sin \left[\frac{\pi r}{r_S} \right] \left[3 \sqrt{z / r} - 1 \right]
\end{equation}
and extract the value of the density at several locations between the center and the equator every few iterations. In post-processing, these
time sequences are again Fourier-transformed to obtain the mode frequencies. In the Cowling approximation, we obtain, for the $l = 2$ $f$-mode,
a value of $\nu({}^2f) = 1884 (40) \, \mathrm{Hz}$, compared to $1890 \, \mathrm{Hz}$ reported in \cite{Gaertig08a}. In full general relativity
we get $\nu({}^2f) = 1601 (30) \, \mathrm{Hz}$, which is in marginal agreement with the result $1586 \, \mathrm{Hz}$ 
from the CFC approximation \cite{Dimmelmeier06a}. 

\subsection{Oscillations of a rapidly rotating neutron star}

As a test case for a rapidly rotating model, we select the BU7 polytrope, which is defined by $\rho_c = 1.28 \times 10^{-3}$, $K = 100$,
$\Gamma = 2$ and a coordinate axis ratio $r_p/r_e = 0.65$. For the fundamental quasi-radial F mode, we select the same form of 
the perturbation function as above, but since the star is strongly rotationally deformed we modify the surface coordinate $r_S$ 
to interpolate between the polar radius and the equatorial radius with an ellipsoidal approximation. Frequencies for this mode
have been reported as $\nu(F) = 2546 \, \mathrm{Hz}$ in the Cowling approximation \cite{Gaertig08a} and $\nu(F) = 1207 \,
\mathrm{Hz}$ in the CFC approximation to general relativity \cite{Dimmelmeier06a}. We obtain $2491 (50) \, \mathrm{Hz}$
and $1204 (40) \, \mathrm{Hz}$ in agreement with these values.

For the axisymmetric $\ell = 2$ $f$-mode, we again impose the same perturbation as in the nonrotating case, but with a $\theta$-dependent
value of $r_S$, and extract mode frequencies of $\nu({}^2f) = 1708 (20) \, \mathrm{Hz}$ in the Cowling approximation, and 
$\nu({}^2f) = 1694 (35) \, \mathrm{Hz}$ in full general relativity. The first value matches with $1703 \, \mathrm{Hz}$ cited from
\cite{Gaertig08a}, whereas the second agrees with the CFC approximation ($1720 \, \mathrm{Hz}$) \cite{Dimmelmeier06a}. 

Since we will be concerned with nonaxisymmetric $f$-modes later, we also investigate the $\ell = 2, |m| = 2$ $f$-modes in this model
and compare to the recent study by Gaertig \& Kokkotas \cite{Gaertig08a} in the Cowling approximation. 
The techniques used here are also used to determine the nonaxisymmetric modes in full general relativity, and therefore
we will describe the method in more detail.

To extract fluid perturbations of a particular $m$, we use a method similar to the one in \cite{Zink2005a, Zink2006a}: On the 
equatorial plane of the star, we specify coordinate radii $r_i$, discretize the corresponding circles with sample points (usually
100 are sufficient), interpolate the density onto the sample points, and project the resulting density angular profiles by discrete
Fourier transforms. The resulting data output every few iterations gives a time profile of the Fourier amplitude of the nonaxisymmetric
part of the density for each mode $|m| \neq 0$ in the analysis. We normalize these perturbations to the average density on the
circle and denote the amplitudes $A_m$ to distinguish between the mode strengths.

This method is particularly useful to identify the $\ell = |m|$ $f$-modes for the following reasons. First, the modes have a dominant
pressure perturbation, which, by the adiabatic condition, also implies a dominant density perturbation. Second, the spherical
harmonics $Y_{l,|l|}$ have meridional node lines $\phi = \mathrm{const.}$ which give orthogonal projections with the corresponding
Fourier basis functions in the equatorial plane. Third, the artificial stellar drift associated with the discretization error
(which appears as low-order components in the post-processing time Fourier transform of the axisymmetric average 
for $m = 0$) does not appear in the profiles for $|m| > 0$.

An additional tool we can use is the choice of a free phase function in the extraction. The mode projection on each circle
is naturally written as a projection onto $\exp (i m (\phi + \phi_0))$, where the phase angle $\phi_0$ can be chosen freely.
In particular, we will use a time-dependent function of the form $\phi_0 (t) = \Omega t$, where $\Omega$ is the
rotational angular frequency of the stellar background model. This choice of phase shifts frequencies by a factor of 
$m \Omega$, which is useful for extraction of mode frequencies near corotation (see also \cite{Gaertig08a}).

\begin{figure}
\includegraphics[width=\columnwidth]{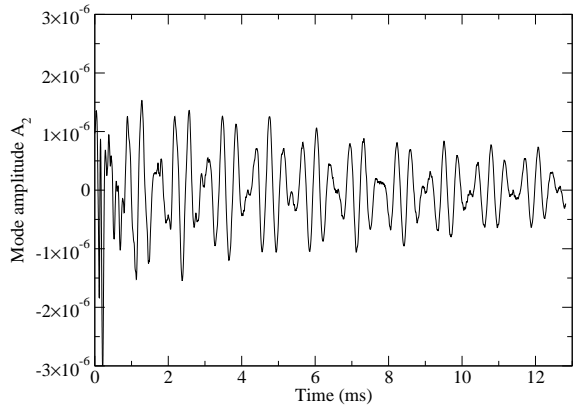}
\caption{Example of a signal obtained by the mode extraction technique described in the text. The background
model is a member of the sequence C (number C1, see Table~\ref{table:c_models}), with $K = 1000$,
$\Gamma = 2.5$ and $\rho_c = 5 \times 10^{-3}$. A nonaxisymmetric $\ell = 2$, $m = 2$ density
perturbation excites the corresponding two $f$-modes, and the resulting oscillations are extracted by
a discrete Fourier analysis of the density on discrete rings. The plot shows the resulting signal of the 
mode $A_2$ (for $m = 2$) for an evolution in full general relativity over $13 \, \mathrm{ms}$. 
The signal exhibits oscillations which are dominated by two frequencies. Also visible is a decay in the 
signal strength, which is a numerical artifact caused by the finite-volume scheme and the
inaccurate discretization near the stellar surface.}
\label{fig:s4t0.02c_signal}
\end{figure}

An example of a signal\footnote{We will frequently refer to the numerically generated time series as a signal
to underline that we are performing numerical experiments.} obtained in this way is given in Fig.~\ref{fig:s4t0.02c_signal}. The particular
model used here is number C1 of the C sequence (see Table~\ref{table:c_models}), with $K = 1000$,
$\Gamma = 2.5$ and $\rho_c = 5 \times 10^{-3}$. The model is evolved, in the Cowling approximation,
for about $13 \, \mathrm{ms}$. After an initial transient,
the two frequencies corresponding to the two $|m| = 2$ $f$-modes are already visible in the signal.
The signal also exhibits a decay over time. Gravitational radiation will reduce the mode amplitude
over time (both $\ell = |m| = 2$ $f$-modes are stable in this model), but on much longer timescales than
in the plot. The decay is therefore a result of numerical dissipation in the finite-volume scheme,
possibly in part also due to inaccuracies of the evolution near the surface. In practice, this decay
is not a concern as long as a sufficiently large number of cycles are available to extract the mode frequency\footnote{Actually,
a damped oscillator has a \emph{different} peak frequency than its undamped counterpart; we 
have estimated this effect from the damping timescale, but have found that it is far below the error bars reported in
this publication.}, i.e. it limits the total time interval $\delta t$ available for mode extraction, and
therefore also the total frequency resolution in post-processing.

There are three direct ways to reduce this effect: (i) Increase the resolution. Since numerical dissipation
scales with the grid spacing, a lower spacing also reduces the mode damping. In addition, the noise
level is slightly lower when going to higher resolutions, and it is in part also determined by the
(fixed) floating point accuracy. Over all, higher spatial resolutions therefore also imply higher
frequency resolutions in post-processing. In practice, three-dimensional simulations in full general relativity are
still resource-intense, and given the systematical uncertainties in the choice of background model
and microphysics we feel that a moderate frequency resolution as used here is sufficient. (ii) Increase
the initial perturbation amplitude. Assuming the same damping, this will allow for a longer extraction
windows and increase the frequency resolution. However, high initial amplitudes
increase the relevance of nonlinear effects, which we do not intend to study here. Experiments show
that this is probably not important until perturbation amplitudes larger than $A = 0.01$ are being
used, but for safety we use $A = 0.001$ exclusively here. (iii) Use a different numerical scheme. Since
 the numerical dissipation is a result of the hydrodynamical scheme (it also appears in evolutions in the
Cowling approximation) a change of scheme has the potential to improve the result. We have experimented
with a monotonized central (MC) reconstruction instead of PPM, and found that with MC the damping
is \emph{less} prevalent. However, MC reconstruction also introduces a stronger angular momentum drift
in the star.

\begin{figure}
\includegraphics[width=\columnwidth]{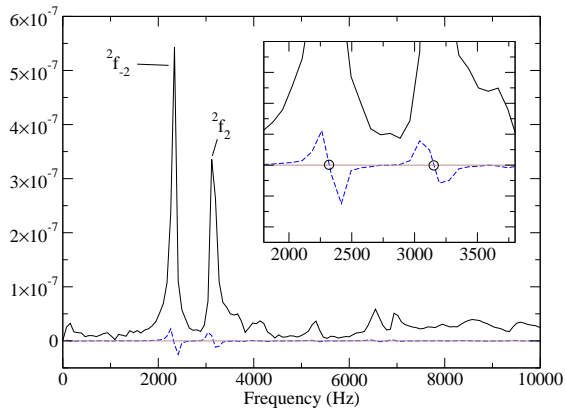}
\caption{Fourier transform of the signal from Fig.~\ref{fig:s4t0.02c_signal} obtained in post-processing.
The initial $l = 2, m = 2$ perturbation excites both the co- and counter-rotating $f$-modes. To determine the
actual  mode frequency, we do not use the location of the peak, but the root of the derivates obtained
from second order central differencing, as shown in the inset.}
\label{fig:s4t0.02c_psd}
\end{figure}

A direct Fourier transform of the mode signal from Fig.~\ref{fig:s4t0.02c_signal} in post-processing leads to the
spectrum shown in Fig.~\ref{fig:s4t0.02c_psd}. Both the co- and the counter-rotating $l = 2$ $f$-modes are excited by the 
perturbation. As explained earlier, the frequencies are obtained from the roots in the derivative of the power spectral density,
which we approximate with second order accurate centered finite differences.

Previous results for nonaxisymmetric modes in rapidly rotating stars have been obtained,
in the Cowling approximation and using linearized equations, by Gaertig \& Kokkotas \cite{Gaertig08a}
and Kr\"uger, Gaertig \& Kokkotas \cite{Krueger09a}. 
To compare with these results we have evolved the rapidly rotating model BU7 
 with our nonlinear code in the Cowling approximation. 
We obtain, for the corotating $\ell = 2$ $f$-mode in the laboratory frame, a frequency of 
$2447 (25) \, \mathrm{Hz}$, compared to $2454 \, \mathrm{Hz}$ in \cite{Gaertig08a}, and for
the counter-rotating mode $183 (30) \, \mathrm{Hz}$, compared to $210 \, \mathrm{Hz}$.

Overall, our code tests show that we can reliably reproduce published $f$-mode frequencies in rapidly rotating stars
both in the Cowling approximation and in dynamical spacetimes.

%% file: results.tex
\section{Results}
\label{sec:results}

We have investigated the $\ell = |m| = 2$ and $\ell = |m| = 3$ $f$-mode frequencies in two sequences
of different stiffness, both in the Cowling approximation and in full general relativity. In
the following two sections, we describe the results of our simulations for each sequence
separately, and conclude this part with a comparison of the spectra in the two sequences.

\subsection{$f$-mode spectrum of sequence S ($\Gamma = 2.0$)}

The first set of polytropes is the S sequence as described in Table~\ref{table:s_models}, which have been
chosen to compare with the approximate location of the neutral point in general relativity by
Stergioulas \& Friedman \cite{Stergioulas98a}. The major surprise in that publication was that
general relativity substantially enhances the CFS instability in comparison to calculations
assuming a stellar structure given by Newtonian physics. However, the techniques used in
\cite{Stergioulas98a} used a truncated gauge to determine the location of the neutral point, 
which should be less accurate particularly in the low-order modes most relevant for
the CFS instability as an observable source. 

\begin{table}
\begin{center}
\begin{tabular}{ccrrr}
Model & $T/|W|$     & $\nu_{rot}$ & $\nu({}^2 f_{2})$ & $\nu({}^2 f_{-2})$ \\
          &   $\times 10^{-2}$  &  Hz            &  Hz &  Hz \\
\hline
S0 & 0.000 & 0 & 2162 (20) & 2162 (20) \\
S1 & 2.000 & 644 & 1130 (35) & 2888 (35) \\
S2 & 4.000 & 889 & 630 (35) & 3056 (30) \\
S3 & 6.000 & 1059 & 215 (40) & 3134 (30) \\
S4 & 8.000 & 1183 & -240 (30) & 3154 (30) \\
\hline
\end{tabular}
\end{center}
\caption{Nonaxisymmetric $\ell = |m| = 2$ $f$-mode frequencies in general relativistic polytropes
in sequence S ($\Gamma = 2.0$, $\rho_c = 2.6812 \times 10^{-3}$, see also Table~\ref{table:s_models}).
The quantity $\nu_{rot}$ is the rotational frequency of the star, $\nu({}^2 f_{2})$ is the frequency
of the counter-rotating $f$-mode, and $\nu({}^2 f_{-2})$ the corresponding frequency of the corotating
mode. The numbers in brackets are approximate error estimates.}
\label{table:s_l2_gr}
\end{table}

The $\ell = |m| = 2$ $f$-mode frequencies in models S0 to S4 have been obtained with the Fourier extraction
methods described in the last section, and the resulting frequencies (in the inertial frame) 
are listed in Table~\ref{table:s_l2_gr}. As expected, the modes are degenerate in the nonrotating
model and then encounter rotational splitting. The corotating modes saturate at some limiting frequency,
here $\approx 3.2 \, \mathrm{kHz}$, whereas the counter-rotating modes have lower frequencies with higher
stellar rotation, since they are dragged in the direction of the stellar rotation.

\begin{table}
\begin{center}
\begin{tabular}{ccrrrrr}
Model & $T/|W|$  & $\nu_{rot}$ & $\nu_C({}^2 f_{2})$ & $\nu_C({}^2 f_{-2})$ & 
	$\Delta({}^2 f_{2})$ & $\Delta({}^2 f_{-2})$ \\
          &   $\times 10^{-2}$  &  Hz            &  Hz &  Hz & Hz & Hz \\
\hline
S0 & 0.000 & 0 & 2445 (35) & 2445 (35) & $13.1\%$ & $13.1\%$ \\
S1 & 2.000 & 644 & 1392 (35) & 3176 (35) & $23.2\%$ & $10.0\%$ \\
S2 & 4.000 & 889 & 873 (40) & 3351 (40) & $38.6\%$ & $9.7\%$ \\
S3 & 6.000 & 1059 & 427 (40) & 3409 (35) & $98.6\%$ & $8.8\%$ \\
S4 & 8.000 & 1183 & -101 (55) & 3382 (35) & $-57.9\%$ & $7.2\%$ \\
\hline
\end{tabular}
\end{center}
\caption{Comparison of the Cowling approximation with full general relativity for the $\ell = |m| = 2$
$f$-modes in the sequence S. The frequencies obtained from the Cowling approximation are
shown in the columns $\nu_C({}^2 f_{2})$ and $\nu_C({}^2 f_{-2})$ (for the counter- and 
corotating modes). The columns $\Delta({}^2 f_{2})$ and $\Delta({}^2 f_{-2})$ display the
relative error of the Cowling approximation, defined as $\Delta = \nu_{C} / \nu_{GR} - 1$.}
\label{table:s_l2_cowling_vs_gr}
\end{table}

\begin{figure}
\includegraphics[width=\columnwidth]{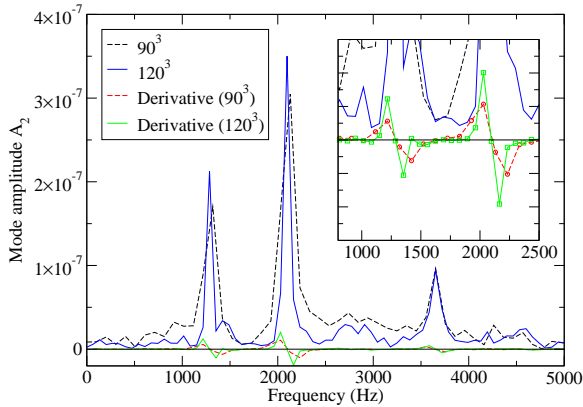}
\caption{Resolution test for the rapidly rotating model S3 (in full GR). Amplitude
extraction results from the Fourier transform in both a simulation with 90 grid cells per coordinate direction
and 120 cells are compared. The finer resolved simulation has also been run for longer time (up to 20 ms)
to reduce peak width and the resulting size of the error bars in the Fourier transform. The graph shows
the Fourier transform, its derivatives and the null line to visually identify the location of the roots. It is apparent
that the discrete peak maximum locations disagree slightly, but the locations of the root agree very well,
better than the error bar estimate obtained from the half-width of the peak.}
\label{fig:resolution_signal}
\end{figure}

To assess the influence of grid resolution on the extracted frequencies, we have performed higher resolution simulations in
selected cases (120
instead of 90 grid cells in each direction) and confirmed that the results are unchanged within the error bars. 
Fig.~\ref{fig:resolution_signal} shows an example of the result for the rapidly rotating model S3.

The Cowling approximation is particularly inaccurate when low-order $f$-modes are considered \cite{Yoshida97a},
and therefore we have also performed the same set of simulations with the spacetime variables kept fixed.
The resulting relative changes in the observed $f$-mode frequencies are detailed in Table~\ref{table:s_l2_cowling_vs_gr}.
Modes obtained from the Cowling approximation generally show an overestimation of the actual frequency.
In the nonrotating model, this amounts to a difference of about $13 \%$ in accord with earlier studies
\cite{Dimmelmeier06a}. For the corotating modes, the relative error in frequency reduces when moving towards
the mass-shedding limit to about $7\%$. However, in the counter-rotating modes, the error grows to rather large
values as the sequence approaches the neutral point (and also beyond that). The definition of the relative error
is clearly singular at the neutral point; therefore, one could view this as a mathematical artifact of the definition (in
particular since the frequencies in the corotating frame will have a smaller error more consistent with the 
corotating modes). However, observationally the difference between $427 \, \mathrm{Hz}$ (Cowling) and 
$215 \, \mathrm{Hz}$ (GR, both model S3) is very relevant, regardless of how the error is defined.

\begin{figure}
\includegraphics[width=\columnwidth]{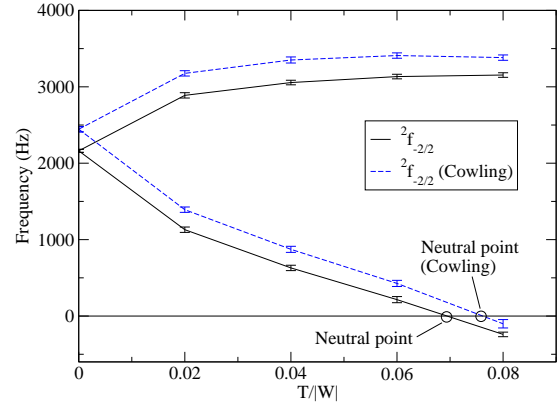}
\caption{Frequencies of $\ell = |m| = 2$ $f$-modes in the models of sequence S ($\Gamma = 2.0$). The
models are characterized by their ratio of rotational kinetic over gravitational binding energy $T/|W|$, where
$T/|W| = 0.08$ is very close to the mass-shedding limit. Both results from simulations in the Cowling
approximation (blue dashed graph) and those from full simulations (black full graph) are displayed for 
comparison. The approximate location of the neutral point is indicated for each graph.}
\label{fig:nu_over_tw_cowling_vs_gr}
\end{figure}

The results from both the full simulations and those using the Cowling approximation are displayed graphically
in Fig.~\ref{fig:nu_over_tw_cowling_vs_gr}. The shift between the Cowling approximation and full simulations
is almost constant over the sequence, which causes the large relative error at low mode frequencies. When
parameterizing the sequence with $T/|W|$ as done here, the frequencies of the counter-rotating $f$-modes are
almost a linear function, which makes a linear interpolation to locate the neutral point more reliable (this is
in contrast to a parameterization over the rotational frequency $\nu$, cf. \cite{Gaertig08a}). In the Cowling approximation,
the neutral point is located at $T/|W|_n = 0.076 \pm 0.002$, whereas in the full simulations we find 
$T/|W|_n = 0.0695 \pm 0.0018$, i.e. the Cowling approximation tends to underestimate the lower limit of the 
$f$-mode CFS instability window by about $10 \%$ in this particular case.

The actual location of the neutral point is of some interest since we can compare it with the results obtained from
perturbation theory. In \cite{Stergioulas98a}, the location of the neutral point is quantified, for this particular
sequence, at $T/|W|_n = 0.0649$ (Table 3 in \cite{Stergioulas98a}). This is outside of the approximate confidence
level we have reported above by $3 \%$, and in disagreement with the average value by $5.9 \%$. The most likely
cause for this disagreement is the use of the truncated gauge in \cite{Stergioulas98a} since it is expected to be
less accurate at lower order. We will provide some evidence for this possibility below. In this sense, the full numerical
simulation seems to indicate a slightly smaller instability window than previously expected on grounds of the neutral
point location.

\begin{table}
\begin{center}
\begin{tabular}{ccrrr}
Model & $T/|W|$     & $\nu_{rot}$ & $\nu({}^3 f_{3})$ & $\nu({}^3 f_{-3})$ \\
          &   $\times 10^{-2}$  &  Hz            &  Hz &  Hz \\
\hline
S0 & 0.000 & 0 & 2758 (30) & 2758 (30) \\
S1 & 2.000 & 644 & 1053 (30) & 4067 (40) \\
S2 & 4.000 & 889 & 229 (30) & 4405 (35) \\
S3 & 6.000 & 1059 & -457 (40) & 4535 (60) \\
S4 & 8.000 & 1183 & -1307 (55) & 4572 (50) \\
\hline
\end{tabular}
\end{center}
\caption{Frequencies of $\ell = |m| = 3$ $f$-modes in the sequence S. Notations as
in table~\ref{table:s_l2_gr}.}
\label{table:s_l3_gr}
\end{table}

\begin{table}
\begin{center}
\begin{tabular}{ccrrrrr}
Model & $T/|W|$  & $\nu_{rot}$ & $\nu_C({}^3 f_{3})$ & $\nu_C({}^3 f_{-3})$ & 
	$\Delta({}^3 f_{3})$ & $\Delta({}^3 f_{-3})$ \\
          &   $\times 10^{-2}$  &  Hz            &  Hz &  Hz & Hz & Hz \\
\hline
S0 & 0.000 & 0 & 2971 (60) & 2971 (60) & $7.7 \%$ & $7.7 \%$ \\
S1 & 2.000 & 644 & 1222 (30) & 4228 (30) & $16.1 \%$ & $4.0 \%$ \\
S2 & 4.000 & 889 & 382 (25) & 4574 (25) & $66.8 \%$ & $3.8 \%$ \\
S3 & 6.000 & 1059 & -331 (30) & 4713 (35) & $-27.6 \%$ & $3.9 \%$ \\
S4 & 8.000 & 1183 & -1232 (40) & 4544 (40) & $-5.7 \%$ & $-0.6 \%$ \\
\hline
\end{tabular}
\end{center}
\caption{Comparison of the Cowling approximation with full general relativity for the $\ell = |m| = 3$
$f$-modes in the sequence S. Notations as in Table~\ref{table:s_l2_cowling_vs_gr}.}
\label{table:s_l3_cowling_vs_gr}
\end{table}

To investigate the influence of the mode order on the  extracted frequencies, we also consider the $\ell = |m| = 3$ $f$-modes in the
sequence S. The resulting frequencies are listed in Table~\ref{table:s_l3_gr}, and the comparison between Cowling results
and full simulations can be found in Table~\ref{table:s_l3_cowling_vs_gr}. The frequencies of the corotating $\ell = 3$ modes
are generally higher, in accord with expectations from perturbation theory: we have $2.8$ to $4.6 \, \mathrm{kHz}$ as 
opposed to $3.2 \, \mathrm{kHz}$ in the $\ell = 2$ case. The Cowling approximation is more accurate for these
modes. 

\begin{figure}
\includegraphics[width=\columnwidth]{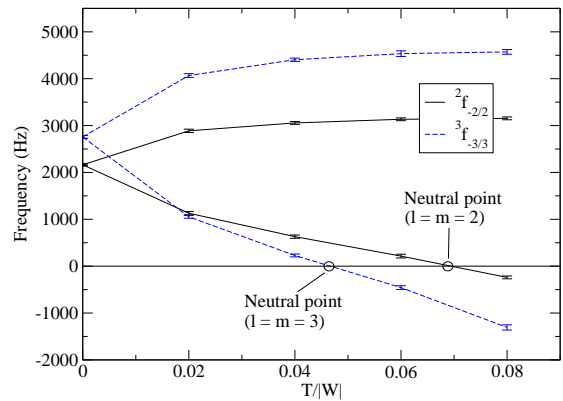}
\caption{Comparison between $\ell = |m| = 2$ and $\ell = |m| = 3$ $f$-mode frequencies along sequence S. The neutral point
of the $m = 3$ mode is shifted with respect to the $m = 2$ mode.}
\label{fig:nu_over_tw_s_m2_vs_m3}
\end{figure}

Fig.~\ref{fig:nu_over_tw_s_m2_vs_m3} shows a comparison between the $\ell = 2$ and $\ell = 3$ frequencies obtained
from the full simulations. The higher order mode becomes unstable at lower \footnote{We remind the reader that
these statements apply to \emph{ideal fluid polytropes}, and that viscosity and superfluidity determine the actual instability
window in real neutron stars. However, the neutral point is a lower limit for this window.} $T/|W|$, here at 
$T/|W|_n = 0.0467 \pm 0.0010$. 

We again compare this value with predictions from linear theory: the neutral point for the $\ell = m = 3$ $f$-mode from
\cite{Stergioulas98a}, Table 3 is stated as $T/|W|_n = 0.0455$, which is only in much better agreement with our
full nonlinear result ($2.6 \%$ when comparing average values). This behavior is expected from the use of 
a truncated gauge in \cite{Stergioulas98a}, and therefore can be cited as evidence identifying this approximation
as the cause of the disagreement. In addition, the errors for the choice of a truncated gauge are estimated in 
\cite{Stergioulas98a} for a model with somewhat smaller central density than used here.  Using the
results of \cite{Stergioulas98a} for the lower-density model (full gauge: $T/|W|_n = 0.0463$, 
truncated gauge: $T/|W|_n = 0.0459$) and assuming that the relative error between the full and
truncated gauge remains roughly the same at the somewhat larger central density used here, 
we find that the disagreement between our current results and those published
in \cite{Stergioulas98a} for the $l=m=3$ neutral points is less than $1.7 \%$, which is within our current error bars.

\subsection{$f$-mode spectrum of sequence C ($\Gamma = 2.5$)}

The sequence C (cf. Table~\ref{table:c_models} is designed to investigate the influence of stiffness in the equation
of state on the nonaxisymmetric $f$-modes and on the location of the neutral point. All models have a central density
which is close to the central density of the maximum mass nonrotating model, similar to the S sequence. The main difference
is that sequence C admits substantially higher rotation rates compared to sequence S: while the mass-shedding
limit lies around $T/|W| \approx 0.09$ in sequence S, sequence C has a limit beyond $T/|W| \approx 0.12$.
This is relevant not only for the range of frequencies expected from the CFS instability, but also for the growth time
of an unstable mode.

\begin{table}
\begin{center}
\begin{tabular}{ccrrr}
Model & $T/|W|$     & $\nu_{rot}$ & $\nu({}^2 f_{2})$ & $\nu({}^2 f_{-2})$ \\
          &   $\times 10^{-2}$  &  Hz            &  Hz &  Hz \\
\hline
C0 & 0.000 & 0 & 2527 (70) & 2527 (70) \\
C1 & 4.000 & 1046 & 698 (50) & 3668 (50) \\
C2 & 8.000 & 1422 & -170 (50) & 3894 (60) \\
C3 & 12.00 & 1654 & -1029 (50) & 3995 (60) \\
\hline
\end{tabular}
\end{center}
\caption{Frequencies of $\ell = |m| = 2$ $f$-modes in the sequence C. Notations as
in table~\ref{table:c_l2_gr}.}
\label{table:c_l2_gr}
\end{table}

We report the frequencies for the $\ell = |m| = 2$ $f$-modes in general relativistic simulations of sequence C in 
Table~\ref{table:c_l2_gr} (note the comment about the choice of units in \ref{sec:initial_models}). The general behavior 
of the frequencies is comparable to the softer equation of state, but the lower compressibility of the material shifts
the corotating mode frequencies into higher bands. In addition, the larger maximum rotation rate introduces a much higher limiting
frequency of the counter-rotating $f$-mode, for which we find a lower limit already in the kHz regime. The differences between
full simulations and the Cowling approximation are not listed here, but they show a similar shift as in the S sequence, which again becomes particularly important at lower frequencies. However, even for the fastest rotator C3 we have investigated, the Cowling result is $-728 (45) \, \mathrm{Hz}$ compared to $-1029 (50) \, \mathrm{Hz}$ reported in Table~\ref{table:c_l2_gr}, resulting in a relative difference of $-29 \%$.

\begin{figure}
\includegraphics[width=\columnwidth]{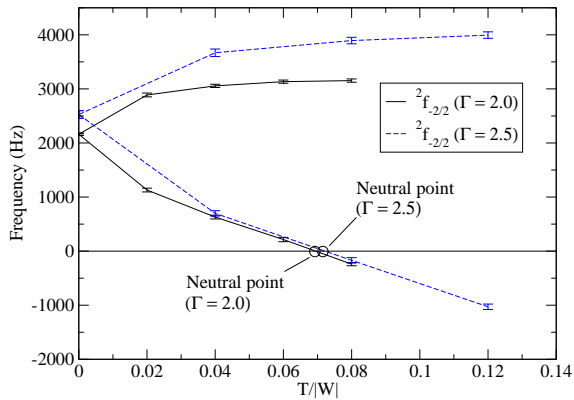}
\caption{Influence of the equation of state on the $\ell = |m| = 2$ $f$-mode frequencies. The results are from
simulations of sequences S and C (cf. Tables~\ref{table:s_models} and \ref{table:c_models}, 
also Fig.~\ref{fig:nu_over_tw_cowling_vs_gr}). While the corotating modes directly show the effect of the
lower compressibility of the $\Gamma = 2.5$ models in a shift in the frequency band, the counter-rotating
modes have comparable frequencies in the admissible range of stellar modes, and consequently a similar
location of the neutral point in terms of T/|W|.}
\label{fig:nu_over_tw_l2_s_vs_c}
\end{figure}

In Fig.~\ref{fig:nu_over_tw_l2_s_vs_c} we compare the mode frequencies between the S and C sequences to illustrate
the influence of the equation of state on the $f$-modes. The less compressible models show larger frequencies in 
the corotating modes, and extend to higher negative frequencies in the counter-rotating modes as observed above. The location
of the neutral point in terms of $T/|W|$ is very similar when comparing these particular sequences. 

\begin{table}
\begin{center}
\begin{tabular}{ccrrr}
Model & $T/|W|$     & $\nu_{rot}$ & $\nu({}^3 f_{3})$ & $\nu({}^3 f_{-3})$ \\
          &   $\times 10^{-2}$  &  Hz            &  Hz &  Hz \\
\hline
C0 & 0.000 & 0 & 3282 (50) & 3282 (50 \\
C1 & 4.000 & 1046 & 332 (40) & 5374 (40) \\
C2 & 8.000 & 1422 & -1104 (40) & 5828 (50) \\
C3 & 12.00 & 1654 & & 5864 (40) \\
\hline
\end{tabular}
\end{center}
\caption{Frequencies of $\ell = |m| = 3$ $f$-modes in the sequence C. Notations as
in Table~\ref{table:c_l2_gr}. We were unable to extract to counter-rotating $f$-mode
frequency in the most rapidly rotating model, therefore it is left open here.}
\label{table:c_l3_gr}
\end{table}

\begin{figure}
\includegraphics[width=\columnwidth]{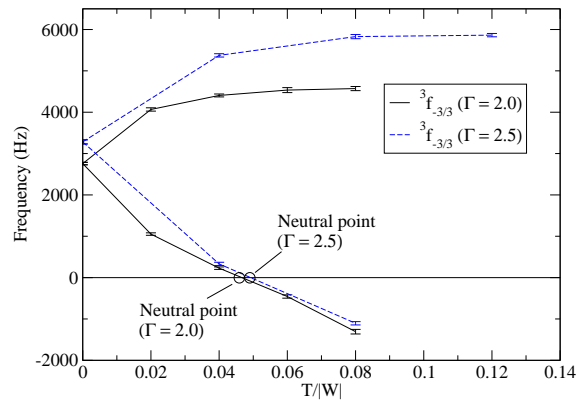}
\caption{Same as Fig.~\ref{fig:nu_over_tw_l2_s_vs_c}, but for the
$\ell = |m| = 3$ modes. }
\label{fig:nu_over_tw_l3_s_vs_c}
\end{figure}

The $\ell = |m| = 3$ $f$-modes in sequence C are listed in Table~\ref{table:c_l3_gr}. We were unable
to extract the frequency of the counter-rotating mode in the most rapidly rotating model C3 due
to numerical noise levels, therefore this information is left out of the table. We display
a comparison to the same modes in the sequence S in Fig.~\ref{fig:nu_over_tw_l3_s_vs_c}, The
neutral point is again only mildly different between these sequences, as are the frequencies of 
the counter-rotating modes.

%% file: conclusion.tex
\section{Conclusion}
\label{sec:conclusion}

We have analyzed the spectrum of $\ell = |m| = 2$ and $\ell = |m| = 3$ $f$-modes in rapidly rotating neutron stars in full general relativity.
As mentioned in the introduction, these modes are interesting not only for general purposes of neutron star asteroseismology,
but particularly because they have a lower frequency (counter-rotating) branch, and since a part of that branch can be 
unstable to the CFS instability. Our study allows us to investigate the observable frequency band covered by the CFS 
instability (if it is active) within the restrictions imposed by using a polytropic model.

\begin{figure}
\includegraphics[width=\columnwidth]{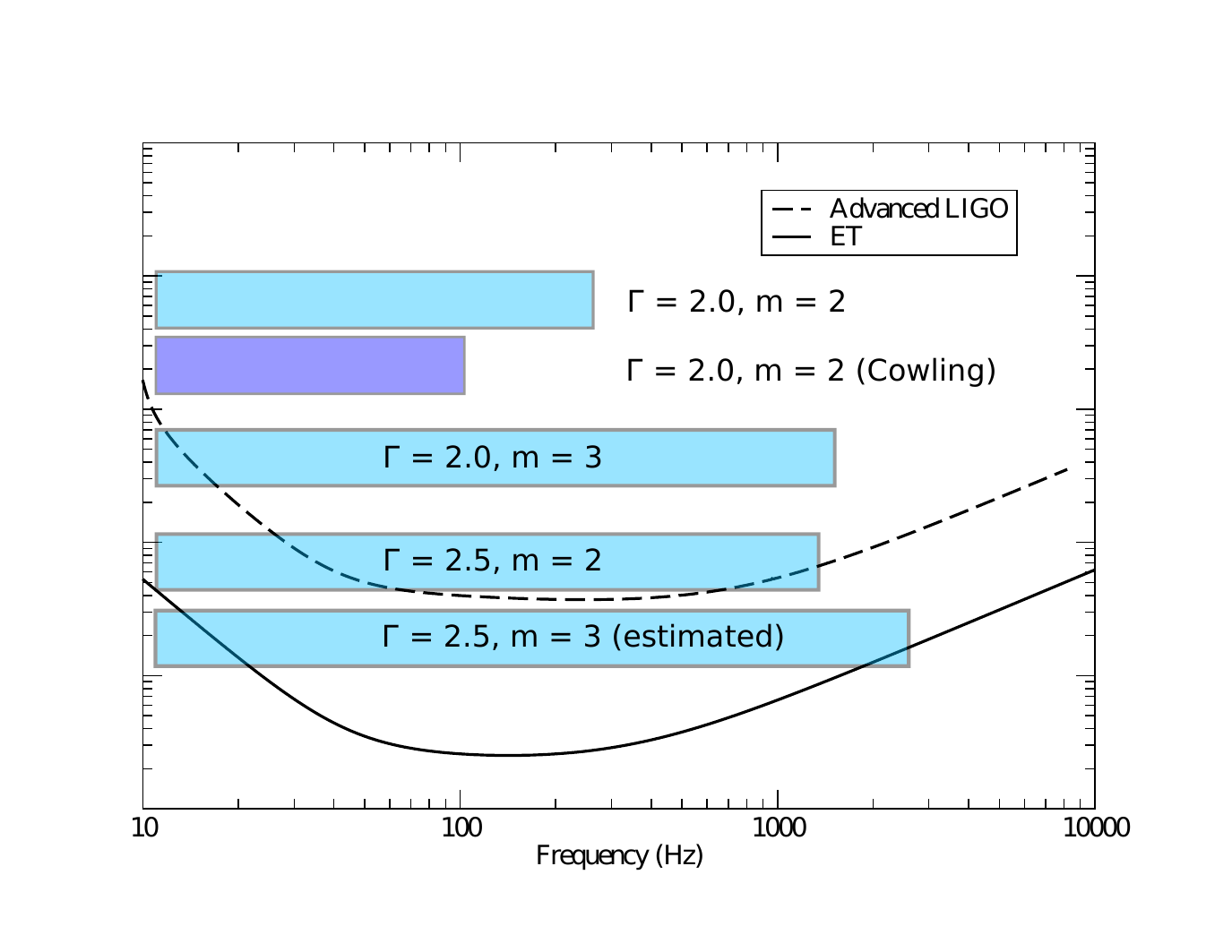}
\caption{Frequency band of the CFS instability based on the extraction of $m = 2$ and
$m = 3$ $f$-mode frequencies in two sequences of uniformly rotating polytropes. Each
band is constrained by the neutral point (low frequencies on the left end) and the frequency
of the counter-rotating mode in the most rapidly rotating model. For the $m = 3$ mode
in the $\Gamma = 2.5$ sequence
of polytropes, the counter-rotating mode frequency is extrapolated from figure~\ref{fig:nu_over_tw_l3_s_vs_c}
to around $\approx 2.5 \, \mathrm{kHz}$ (a lower limit is given by $1.1 \, \mathrm{kHz}$). To compare with
the frequency variation of the detector sensitivity, the noise curves (in $1/\sqrt{\mathrm{Hz}}$) of Advanced LIGO and the proposed 
Einstein Telescope are also displayed.}
\label{fig:frequency_bands}
\end{figure}

The location of the frequency bands available to the CFS $f$-mode instability is displayed in Fig.~\ref{fig:frequency_bands}. 
The diagram also shows the noise curves of Advanced LIGO and the proposed Einstein Telescope to compare with the characteristic
frequency dependence of the detector sensitivity. Actual strain amplitudes of the CFS instability can of course not be determined with the results of this study, but it is instructive to see how the variation of the mode order and the stiffness in the polytropic EOS affects the
available frequencies. 

The band is determined by the location of the neutral point (where the pattern speed introduces very low frequency gravitational waves)
and the frequency of the counter-rotating mode of the fastest rotator. If we take the approximate development of the instability
as described by Lai and Shapiro (see figures 4 and 6 in \cite{Lai95}) at face value, we expect the signal of an unstable star to
rise to high amplitudes with frequencies close to the nonaxisymmetric mode which caused the instability and then sweep through
the band down near the neutral point to a stable state in which the star has lost its angular momentum due to gravitational radiation.
However, since the modes have to compete with viscosity, and since the growth time is shorter for stars with rotation rates 
which are farther removed (for that particular mode) from the neutral point, it appears more likely that models near the breakup limit
are relevant for observation than those near the neutral point. If that is true one would expect the full band displayed here to be 
a fair assessment of the actual observable frequency sweep \emph{in these particular models}. It goes without saying that the actual
band also depends on the particulars of different polytropic sequences (different central density), scales trivially with the polytrope
mass range, and will be different when more realistic stellar models are assumed. However, since the low-order $f$-modes mostly depend on
average density and compressibility of the star, it is quite possible that the systematic scalings observed in these bands readily transfer to 
more realistic models.

From Fig.~\ref{fig:frequency_bands} we can see that the upper third of the $m = 2$
frequency band of the $\Gamma = 2.0$ polytropes investigated by Stergioulas and Friedman \cite{Stergioulas98a} 
lies in the optimal sensitivity range of both Advanced LIGO and the Einsteine Telescope. As explained above, this 
part of the band is also expected to be the most relevant one for purposes of detecting the CFS instability. We can also
see from the diagram that the Cowling approximation is particularly inaccurate in this case, since it underestimates
the upper band by over $60 \%$ ($\approx 100 \, \mathrm{Hz}$ in model S4 as opposed to the full evolution value 
of $\approx 240 \, \mathrm{Hz}$). This disagreement, however, diminishes in sequences with a larger ratio of the Kepler frequency
to the rotational frequency of the neutral point.

When increasing the mode order to $m = 3$, the band extends to about $1.3 \, \mathrm{kHz}$, such that the most likely
observations (from the larger growth time of the instability farther from the neutral point as well as the larger number of cycles,
see \cite{Lai95}) will occur in a band around or below $1 \, \mathrm{kHz}$. A similar change occurs when changing the sequence
to a much stiffer polytrope with $\Gamma = 2.5$, but retaining the mode order to quadrupole. While the emission of the most
unstable polytropes is not anymore in the optimal regime of around $100 - 200 \, \mathrm{Hz}$ it is still expected below
$1 \, \mathrm{kHz}$. Finally, the full extent of the $m = 3$ band in the $\Gamma = 2.5$ sequence could not 
be deduced from our data due to numerical problems, we know a lower limit of around $1.1 \, \mathrm{kHz}$ and, by extrapolation
in the mostly linear part of the $\nu(T/|W|)$ diagram (cf. Fig.~\ref{fig:nu_over_tw_l3_s_vs_c}), should be around $2.5 \, \mathrm{kHz}$.
From earlier work on the location of the neutral point when changing the central density of the star \cite{Stergioulas98a} it is
expected that sequences whose nonrotating members are further from the fundamental mode stability limit $dM/d\rho_c = 0$ 
should have a neutral point at higher ratios of the critical to maximal rotational frequency, and therefore will be more stable to the CFS
mechanism, and would also occupy a smaller frequency band of instability.

With the present study, we have investigated the actual mode frequencies of particular nonaxisymmetric modes in uniformly
rotating, general relativistic polytropes, and were able to infer information about the general location of the frequencies 
the instability could cover. There are a number of important directions for future work:
The first one concerns the rotational profile of the star, since differential rotation opens up a larger instability window by the simple
fact that it admits $T/|W|$ beyond the Kepler limit for uniformly rotating stars, and it is expected that proto-neutron stars are
differentially rotating. For recent work on nonaxisymmetric modes in rapidly differentially rotating stars in the Cowling approximation
see \cite{Krueger09a}. The second limitation is the restriction to a polytropic stratification. A number of equations of state for
cold neutron stars are available and can be used to more closely approximate the structure of a young neutron star. Alternatively,
actual structures of proto-neutron stars obtained from core-collapse simulations could serve as a guide to construct more realistic
models. The third uncertainty concerns the role of superfluidity in modifying the mode spectrum (and thus both the location of the
neutral point and the frequency band). In addition, mutual friction in a superfluid core may modify the instability window in older
neutron stars or even suppress the instability entirely. Related to this, it is important to gain a better understanding of the role
of bulk and shear viscosity in affecting the instability window and the frequency band of the $f$-mode CFS instability. 
Finally, it is important to understand the nonlinear couplings that will lead to a saturation of the mode amplitude as well
as the overall nonlinear development of instability, including the time-evolution of the expected gravitational-wave signal.